\documentclass{article}
\usepackage{ijcai19}
\usepackage{times}
\usepackage{soul}
\usepackage{url}
\usepackage{hyperref}
\hypersetup{hidelinks}
\usepackage[utf8]{inputenc}
\usepackage[small]{caption}
\usepackage{graphicx}
\usepackage{amsmath}
\usepackage{booktabs}
\usepackage{xcolor}
\usepackage{comment}
\usepackage{enumitem}
\usepackage[noend]{algpseudocode}
\usepackage{algorithmicx,algorithm}
\title{Trimming the Sail: A Second-order Learning Paradigm for Stock Prediction}

\author{
Chi Chen$^1$\and
Li Zhao$^2$\and
Wei Cao$^2$\and
Jiang Bian$^2$\And
Chunxiao Xing$^1$\\
\affiliations
$^1$Tsinghua University\\
$^2$Microsoft Research\\
\emails
chenchi14@mails.tsinghua.edu.cn
\{lizo, Jiang.Bian, Wei.Cao\}@example.com,
xingcx@tsinghua.edu.cn
}

\hypersetup{draft}
\begin{document}

\maketitle

\begin{abstract}
Nowadays, machine learning methods have been widely used in stock prediction. Traditional approaches assume an identical data distribution, under which a learned model on the training data is fixed and applied directly in the test data. Although such assumption has made traditional machine learning techniques succeed in many real-world tasks, the highly dynamic nature of the stock market invalidates the strict assumption in stock prediction. To address this challenge, we propose the second-order identical distribution assumption, where the data distribution is assumed to be fluctuating over time with certain patterns. Based on such assumption, we develop a second-order learning paradigm with multi-scale patterns. Extensive experiments on real-world Chinese stock data demonstrate the effectiveness of our second-order learning paradigm in stock prediction.
\end{abstract}

\section{Introduction}

Stock prediction, with the aim at predicting future price trend of stocks, is one of the most important fundamental techniques for stock investment~\cite{preethi2012stock}.
To facilitate stock prediction, traditional quantitative investment approaches usually recognize some trading indicators and then conduct predictions based on these indicators~\cite{suh2004novel}. 
%Analogous to feature engineering in the general machine learning approach, technical trading indicators are usually hand-crafted by professional financial analysts and are developed for providing reasonable and reliable information for stock prediction. 
%For instances, there are many widely-used trading indicators, including moving averages, RSI, and MACD etc. 
%Based on these indicators, financial analysts can discover practical patterns, such as well-known head and shoulders \cite{osler1995head} and cup and handle \cite{suh2004novel}, for accurate stock prediction.
Recently, substantial machine learning techniques have been introduced into stock prediction, since its strong capability in automatically identifying underlying patterns over indicators from the historical data with little human knowledge \cite{patel2015predicting,cervello2015stock}.

Formally, a typical machine learning approach intends to learn a parameterized function \(F_\theta\), mapping the input features \(X\), i.e., various trading indicators, into the output target \(Y\), i.e., the stock future trend. 
While recent years have witnessed a variety of machine learning techniques with different forms of $F_\theta$, such as Linear Regression \cite{zhang2014causal}, Random Forest \cite{khaidem2016predicting}, Neural Networks ~\cite{zhang2017stock,nelson2017stock,fischer2018deep}, etc., 
%have been explored and achieve remarkable performance in stock prediction. 
typical learning-based approaches for stock prediction feel pain when facing the dynamic nature of the stock market.

\begin{comment}

\begin{figure}[!tp]
\centering
%\setlength{\abovecaptionskip}{0.1cm}
%\setlength{\belowcaptionskip}{-0.5cm}
\includegraphics[width=0.45\textwidth]{figs/data_model_evolve.png}
\caption{The optimal model of the stock market evolves through time. Both the underlying data distribution $P_t(X,Y)$ and the optimal prediction model $F_{\theta_t}(X)$ are changing over time. 
Unlike traditional machine learning tasks which have stationary data distribution, we need to consider the highly dynamic nature when dealing with stock market modeling.}
\label{fig:model_evolve}
\end{figure}
\end{comment}
\begin{figure}[!tp]
\centering
\includegraphics[width=0.5\textwidth]{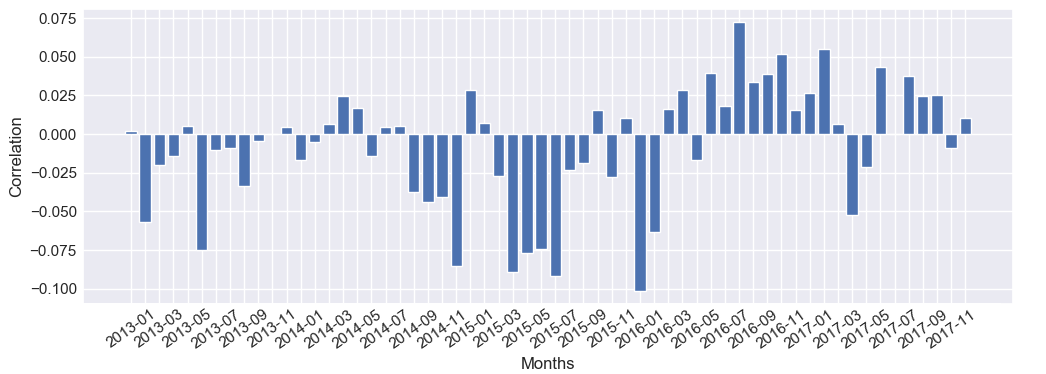}
\caption{{The correlation between the market value of stocks and returns in each month of Chinese market from 2013 to 2017.}}
\label{market_dynamic}
\end{figure}

%traditional machine learning -> (data distribution not change) -> first-order identical distribution assumption -> (stock market invalid and example) -> necessary to consider the changing data distribution
%rotation learning paradigm tracks -> (but cannot capture the sudden altering) -> altering is not unpredictable, some books, rotation market states -> second-order identical distribution assumption -> firstand second-order relation -> our proposed model.....->contribution....
%what is the relation between the market states and data distribution???
Specifically, traditional machine learning approaches usually assume an {\bf i}dentical data {\bf d}istribution (i.d.) $P(X,Y)$. 
{{Thus, after obtaining the optimal $F_{\theta}$ on the training data, the corresponding parameters are fixed and applied directly in the test data.}}
%{\color{red} Thus, the optimal model $F_{\theta}$ learned from the training data can predict over arbitrary future datasets.}
We refer to such assumption as the {\bf first-order i.d. assumption}. Unfortunately, due to the highly dynamic nature of stock market, the data distribution $P(X, Y)$ usually varies over time $t$. 
%We consider the trading indicator ``market value'' as an example.
{{Figure ~\ref{market_dynamic} shows the correlations between the market values of stocks and returns in different months of Chinese market. As we can see, the market value is negatively correlated with the future return before the year of 2016, while positively after 2016.}}
%{\color{blue}{As a result, applying a fixed model trained on a certain period of the historical data does not guarantee the model performance on the future.}}
Thus, it is hard to apply a fixed model to achieve accurate prediction on before and after 2016 simultaneously. 
%cannot guarantee the performance all the time.
In other words, the optimal \textbf{first-order model {\boldmath$F_{\theta}$}} can shift drastically along with different time periods. Therefore, it is fairly important to consider the change of data distribution over time in stock prediction task.

% Since the optimal stock prediction model $F_{\theta^*$ can shift drastically along with different time periods, it is essential to consider the prediction model $F_{\theta_t}$ with different time $t$.
%Therefore, applying a single model trained on a certain set of historical data does not guarantee the performance on future data.
% , since the distribution of the training data may be quite different from that of the test one.

%{\color{blue}{a straightforward and widely used method to address the data distribution variation}}
{{To seek sustaining accurate stock prediction under the critical challenge of non-identical data distribution, a straightforward method}} is to employ the {\em rotation learning paradigm}, which keeps updating new models \(F_{\theta_{[t-\Delta,t]}}\) by rotating the training procedure using merely the most recent data within the certain time window $[t-\Delta,t]$. Nevertheless, the rotation learning paradigm still suffers from a couple of disadvantages. 
%First, since the earlier historical data that are already outside the rotating time window will no longer be used in the training process, it leads to quite low data utilization. Moreover, as it still presumes the same importance for all the data samples within the rotating time window, this learning paradigm overlooks the potential difference of these data samples in terms of their effects on the future inference. 
The most important one is that, even though the rotation learning paradigm has attempted to bridge the gap in terms of the data distribution between the training and the testing data, 
{{it cannot handle sudden distribution altering.}}
%{{\color{red}it cannot recognize the distribution altering immediately}}. 
On the other hand, the distribution variation of financial market is not completely intractable. 
%{\color{blue}{Instead, it usually demonstrates some certain patterns over time. For example, in the famous {\em Merrill Lynch Investment Clock} report \cite{lynch2004investment},  the finance experts claim that the market returns vary over a time loop of four stages.}}
{{Many studies have demonstrated some variation patterns on the financial market. For example, the famous report {\em Merrill Lynch Investment Clock}~\cite{lynch2004investment} claims that the market returns vary over a time loop. Numerous theories of economic cycle have been proposed by many financial professors~\cite{lucas1980methods,choe1993common,naes2011stock}.}}
%N{\ae}s et al. propose business cycle is important to the market liquidity~\cite{naes2011stock}.
Motivated by this, we propose second-order i.d. assumption.

\begin{table}[!tp]
\centering
\caption{Trading indicators and their categories with respective calculation formulas, where \(p_{open}(t)\), \(p_{close}(t)\), \(p_{high}(t)\) and \(p_{low}(t)\) denote the opening price, closing price, highest price and lowest price at time \(t\), and \(m\) is the size of sliding time window.}
\label{tenchnical_indicators}
\begin{tabular}{|l|p{0.75\linewidth}|}
\hline
\multicolumn{1}{|c|}{\textit{\small{Indicators}}} & \multicolumn{1}{c|}{\textit{\small{Calculation Formula}}}

	\tabularnewline \hline
    \tiny{$KLEN(t)$} & \tiny{$(p_{close}(t)- p_{open}(t))/(p_{open}(t))$}
    \tabularnewline \hline
    \tiny{$KUP(t)$} & \tiny{$[p_{high}(t) - \max(p_{open}(t), p_{close}(t))]/p_{open}(t)$}
    \tabularnewline \hline
    \tiny{$KLOW(t)$} & \tiny{$[\min(p_{open}(t), p_{close}(t)) - p_{low}(t)]/p_{open}(t)$}
    \tabularnewline \hline
    \tiny{$MA_m(t)$} & \tiny{$\frac{1}{m}\sum_{j=0}^{j=m-1}p_{close}(t-j)$}
    \tabularnewline \hline
    \tiny{$EMA_m(t)$} & \tiny{$[p_{close}(t)-EMA_m(t-1)]\times \frac{2}{m+1}+EMA_m(t-1)$}
    \tabularnewline \hline
    \tiny{$Bias_m(t)$} & \tiny{$p_{close}(t) - \frac{1}{m}\sum_{j=0}^{j=m-1} p_{close}(t-j)$}
     \tabularnewline \hline
    \tiny{$ROC_m(t)$} & \tiny{$(p_{close}(t) - p_{close}(t-m)) / p_{close}(t-m)$}
    \tabularnewline \hline
    
\end{tabular}
\end{table}
\begin{itemize}[leftmargin=*]
\item
    \textit{\textbf{Second-order i.d. assumption}}. We assume that the data distribution $P(X,Y)$ is fluctuating over time with certain patterns. That is, for each time period $t$, the optimal parameter $\theta_t$ of $F_{\theta_t}$ can be modeled by a \textbf{second-order model} {\boldmath$G$}. Formally,
    %$\theta_t = G(\theta_{t-k},\theta_{t-k+1}, \cdots, \theta_{t-1})$ 
    $\theta_t = G(\theta_{<t})$.
    
    % the prediction model can be predicted by the second-order model $G$ rather than the learned \textbf{first-order model} $F$ from history in the first-order identical distribution assumption.
\end{itemize}

Note that the first-order method can be seen as a special case under second-order  i.d. assumption when the mapping $G$ is an identity function. Based on the second-order i.d. assumption, we propose a novel learning paradigm which attempts to learn the model $G$ from history, and thus derive the proper first-order model to predict the future stock trends more accurately.
%In our proposed framework, the estimated parameter $\theta_t$ from data is regarded as the input of the second-order model $G(\theta_{<t})$. Then, the second-order patterns are modeled by the second-order model $G$.
%,
%{\textcolor{blue}{Note that, the learning target of the second-order learning paradigm is not minimizing the loss of the predicted parameters but the prediction loss of the stock future trend by applying the predicted parameters on the training data.}}
%which minimizes the prediction loss of the stock future trend instead of directly parameters.
%Moreover, to further enhance such second-order learning paradigm, we also take advantage of the multi-scale second-order patterns because different time scales can provide diverse signals to boost the accuracy of the stock prediction. 
% To evaluate the effectiveness of our proposed framework, we conduct extensive experiments on the real-world data in Chinese A-share market. Experiment results demonstrate that our method significantly outperforms the methods which are based on both first-order assumption and rotation learning paradigm.

Our contributions in this paper can be summarized as:
\begin{itemize}[leftmargin=*]
\setlength{\itemsep}{0pt}
\setlength{\parsep}{0pt}
\setlength{\parskip}{0pt}
    \item We identify the first-order i.d. assumption in typical machine learning tasks, which is invalid in stock prediction due to the highly dynamic nature of stock market.
    \item We introduce the second-order i.d. assumption and propose a novel learning paradigm which is able to capture the dynamics of stock market for more accurate prediction.
    \item We conduct extensive experiments on Chinese stock market for more than 2000 stocks over 5 years. Empirical results demonstrate that our paradigm significantly outperforms the first-order methods as well as the rotation learning methods in the stock prediction.
    %Thus, our proposed paradigm can predict the future stock trends more accurately.
\end{itemize}

%The rest of this paper is organized as follows. Section 2 introduces the first-order identical distribution and rotation learning as background. Then, Section 3 proposes a multi-scale sequential learning model which can capture the second-order evolving pattern as introduced. We describe the experimental setup and present experimental results in Section 4. In Section 5, we survey related papers. Finally,we summarize the work in Section 6.

% {{Figure ~\ref{second_model} presents the framework of second-order learning paradigm which contains two major parts: input generation and multi-scale second-order sequential model. In order to capture the second-order patterns, it is intuitive to collect the parameters $\theta$ of the first-order model $F_{\theta}$ first. Then, the second-order patterns based on the generated input are learned by the second-order sequential model. To further enhance such second-order learning paradigm, we also take advantage of the multi-scale second-order patterns because different time scales can provide diverse signals to boost the accuracy of the stock prediction. For adaptive learning, we design the second-order sequential model in a to-end fashion.
% %Finally, the multi-scale outputs are combined for stock prediction. 
% In particular, the target of the second-order learning is to optimize the prediction loss of stock future trends by applying the learned parameters on the training data, instead of the generated parameters.
% }}

The rest of the paper is organized as follows. We first present several preliminaries in Section \ref{sec:pre}. Then, in Section 3, we present our second-order learning paradigm in details. Finally, we demonstrate the experiment results, related work and conclusion in Section 4, 5 and 6.

% the learning target of the second-order learning paradigm is not minimizing the loss of the predicted parameters but the prediction loss of the stock future trend by applying the predicted parameters 
\section {Preliminary}
\label{sec:pre}
%Due to dynamic market, financial experts design and introduce technical indicators, representing patterns of a stock price and volume of diverse aspects, to achieve robust stock prediction~\cite{savin2006predictive,kamijo1990stock,brock1992simple}.
\subsection{Trading Indicator}

Substantial  previous works use trading indicators as the input 
%Recall that the input $X$ of the first-order model $F$ indicates various trading indicators
$X$ of the first-order model $F$~\cite{savin2006predictive,kamijo1990stock,brock1992simple}.
Table \ref{tenchnical_indicators} shows some popular indicators with their respective calculation formulas. Different indicators reflect distinct aspects of trading patterns. Candlestick indicators, such as ``KLEN'', tend to represent trading patterns over short periods of time, usually a few days or a few trading sessions. Trend indicators, such as ``MA'', measure the direction and strength of a trend, using some forms of price averaging. Momentum indicators, such as ``ROC'', identify the speed of price movement by comparing the current closing price to the previous closes.

\subsection{Indicator Effectiveness}
\begin{figure}[!tp]
\centering
\includegraphics[width=0.5\textwidth]{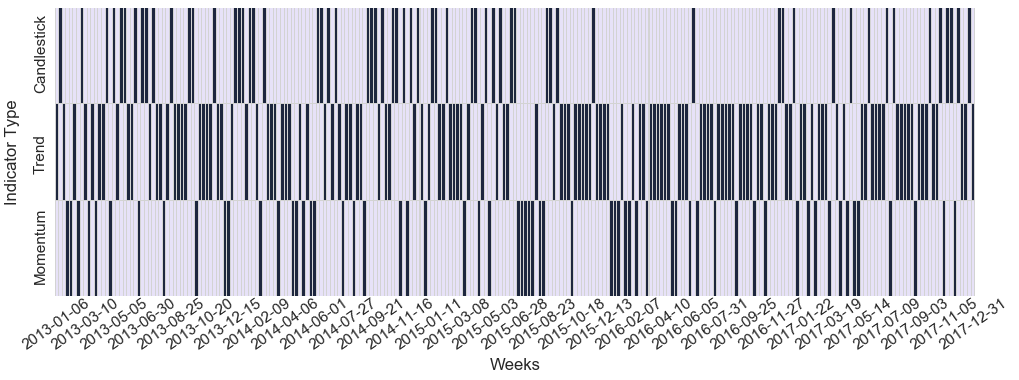}
\caption{The most effective indicator among the three types (``candlestick'', ``trend'' and ``momentum'') in weeks from 2013 to 2017. Each column corresponds to a week, and each row corresponds to a type of indicators. The dark color represents the highest IC value, which means the corresponding type of indicator is the most effective for the stock prediction.}
\label{IC}
\end{figure}
In the financial field, experts usually evaluate the effectiveness of indicators by Information Coefficient (IC)~\footnote{https://en.wikipedia.org/wiki/Information\_coefficient}. The indicator effectiveness reflects the state of the current market. More effective indicators can guide us to a more accurate prediction. Existing first-order methods assume that the effectiveness of indicators stays constant. Thus, once the model finishes training, the corresponding parameters will be fixedly used on any future data.
However, as we mentioned, due to the highly dynamic nature of the stock market, indicator effectiveness is changing over time. Figure ~\ref{IC} shows the change of effectiveness of three types of trading indicators from 2013 to 2017. As we can see, the most effective type does not stay constant but frequent altering, which limits the performance of first-order methods and rotation learning methods. In general, the momentum indicator demonstrates cyclic effective. The trend indicator tends to be more effective while the candlestick indicator is less after the year of 2016. There could be much more complicated patterns of the effectiveness variation which cannot be apparently observed. Therefore, we resort to discover such patterns automatically with a second-order learning paradigm.

\section{Second-order Learning Paradigm}
\begin{figure*}[!t]
\centering
\includegraphics[width=0.95\textwidth]{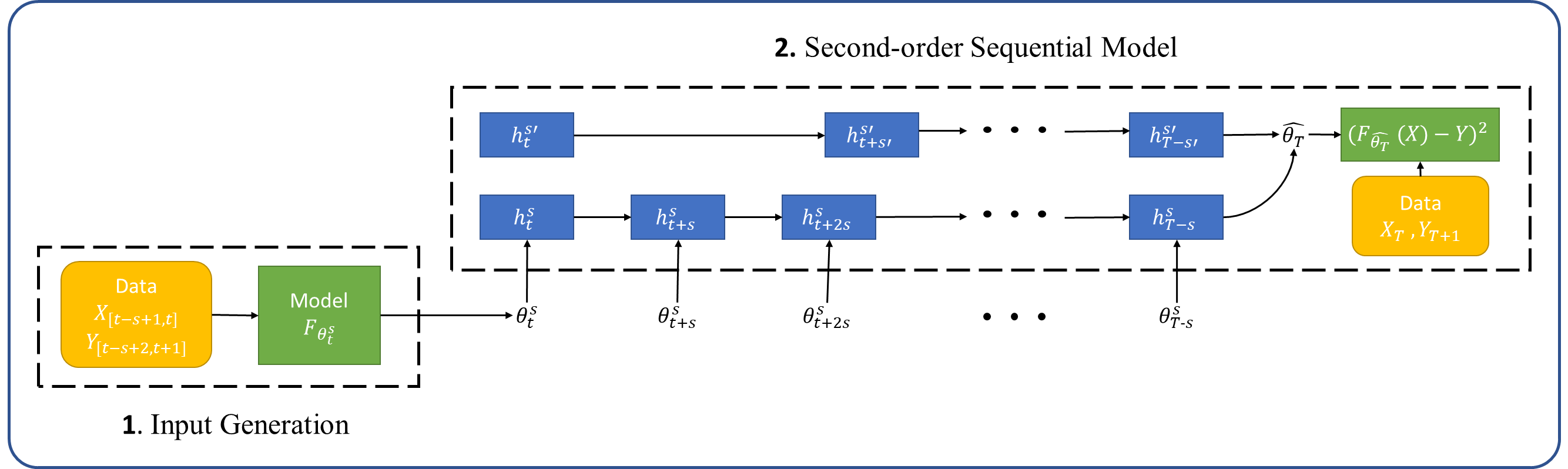}
\caption{The framework of second-order learning paradigm. Our framework consists of two parts. 1. Input Generation: learning parameter $\theta_t$ for prediction model $F_{\theta_t}(X)$ from historical data, as the input of the second part. 2. Second-order Sequential Model: predicting optimal model parameter $\theta_T$ at time step $T$ with diverse time scale in a unified fashion.}
%1. learning parameter $\theta_t$ for prediction model $F_{\theta_t}(X)$ from historical data, as representation of the prediction model. 2. A multi-scale sequential second-order model, predicting optimal model parameter $\theta_T$ at time step $T$. 3. Loss function based on output $\theta_T$.}
\label{second_model}
\end{figure*}
%Stock prediction, with the aim of predicting the future stock trend, plays an important role in the investment. In a more formal way, given data $(X,Y)$, where $X$ represents the trading indicators of the previous days and Y represents the future stock price, our objective is to predict $Y$ based on the input $X$ accurately. Recall that the first-order methods fit a certain parameterized function $Y = F_{\theta}(X)$. 
In this section, we describe our second-order learning paradigm in detail. Our proposed paradigm contains two parts. %To capture the evolving pattern of stock market over time, 
In the first part, we partition the historical data into several periods with multiple time scales. Then, for each time period $t$, we obtain the optimal parameters for the corresponding first-order model $F$. The obtained parameter sequence is used as the input of the second part. Next, in the second part, we use a second-order sequential model $G$ to learn how the optimal parameters $\theta_t$ varies over time using model $G$, and thus predict the future stock trend.

\subsection{Input Generation}
We first present how to generate parameter $\theta_t$, i.e., the input of our second-order model. To capture the evolving patterns of stock market, for each time period $t$, we train a first-order model $F_{\theta_t}$ which generalizes the market state at time $t$. 
% In practice, it is hard to obtain the optimal model. Instead, we use the empirically estimated one, i.e., ${F}_{\hat{\theta}_t}$.
Despite there are many potential types of parameterized function $F_{\theta}$, in this paper, we focus on the linear model because: (1) The data quantity during a small time period is very limited. Thus, the non-linear models such as Decision Tree or Neural Network are prone to overfit the data. (2) For the linear model, each parameter has a well-defined economic meaning.
% as the parameterized function $F_{\theta}$, since it is simple, highly interpretable, and widely used in the financial field.
A linear model can be written as 
\begin{equation}
    F_{\theta}(X) = \mathbf{w}X + b,
\end{equation}
where $\theta=(\mathbf{w},b)$. A positive/negative value of weight $w_i$ implies that $X_i$ yields a positive/negative correlation with the stock trend.  In the meantime, a larger absolute value of $w_i$ usually indicates a more effective trading indicator $X_i$\footnote{However, note that $w_i$ does not directly imply the IC of $X_i$ since we have to consider the co-linearity of the model.}.  Such interpretability is very critical in the financial domain and helps people understand the market dynamics.

% , and the bias $b$ represents the trend of the market. Therefore, the second-order model $G$ learns the change of effectiveness and the market trend in terms of linear first-order model. Such interpretability is critical in the financial domain, which enables experts to understand the underlying patterns discovered by the model. It is worth noting that our proposed framework is not limited to the linear model, but can be applied to any prediction model which is derivable with respect to $\theta$.
In this paper, we actually partition the historical data with multiple time scales. Then, the parameters can be obtained by training the model for each time scale. For the $t$-th time period under time scale $s$, we obtain the parameter vector \(\theta_t^s\) using the historical data \(X_{[t-s+1,t]}\) and the corresponding label \(Y_{[t-s+2,t+1]}\). Intuitively, the sequence of parameters in macro-scale reflects the long-term trend of market state, while micro-scale reflects the short-term trend. 
% On the other hand, the different scale of ${\theta}$ contains distinct information: the macro-scale one reflects long-term market state, while the micro-scale one corresponds to the short-term state. Multi-scale $\theta$s provide various information and are beneficial to the stock prediction. Therefore, we learn the optimal first-order models at multiple time scales. Specifically, for a time scale $s$, \(\theta_t^s = (w_t^s, b_t^s)\) is trained with the historical data \(X_{[t-s+1,t]}\) and corresponding trend \(Y_{[t-s+2,t+1]}\).

\subsection{Second-order Sequential Model}
In the second part, we propose a second-order sequential model to learn the evolving market trends and predict future stock prices. Due to the temporal dynamics in the stock market, we take advantages of the LSTM modeling ~\cite{hochreiter1997long}, which has been widely used to capture the temporal dependencies in the input sequences.
%Since first-order models $F_{\theta_t}$ generalize the stock market at time $t$, 
%Recall that the empirically learned parameters $\theta_{<t}$ are regarded as the inputs of the second-order sequential model. 
%In particular, we take advantage of the LSTM modeling ~\cite{gers2002learning}, which has been wildly used to construct sequential prediction in many applications.
%LSTM is essentially a special type of Recurrent Neural Networks(RNN)~\cite{medsker2001recurrent} that use hidden states(memory) to model sequential patterns of input data. A definite advantage of LSTM over vanilla RNN is its introducing of ``forget'' gates to store long-term memory and avoid vanishing gradients~\cite{gers2002learning}.
In our case, recall that we obtained multiple parameter sequences with different time scales in the first part. For the $t$-th period under the time scale $s$, we have that
% Furthermore, due to the distinct meaning of the multi-scale $\theta$, their evolving patterns may be completely different. For example, the macro-scale tends to be increasing in bull market while the micro-scale may fluctuate. Therefore, we respectively apply the LSTM to capture the second-order patterns on each scale. More formally,
\begin{equation}
    %h_{t-s}^s = LSTM(\theta_{<t}^s)
    h_{t}^s = \mathrm{LSTM}(h_{t-s}^s, \theta_{t}^s),
\end{equation}
where $h_t^s$ is the corresponding ``hidden vector'' which represents the temporal patterns before $t$. For different time scales, since the macro and micro scales indicate different market trends, we use the attention mechanism to combine the hidden states of different time scales, i.e., 

% Then, we combine the long-term and short-term second-order patterns together to predict new prediction model $\hat{\theta}_{T}$: 
\begin{equation}
\hat{\theta}_{T} = \sum_{s} \chi(\alpha^s  h_{T-s}),
\end{equation}
where 
% \(h_{T-s}^s\) is the hidden state of the last LSTM cell in \(s\)-day scale.
$\chi$ is a fully-connected layer transforming the hidden vector to the predicted parameter, and \(\alpha^s \in R\) is the attention weight of the time scale $s$ which is automatically learned by the model. The output $\hat{\theta}_{T}$ corresponds to the first-order parameter estimated by the second-order sequential model at the future period $T$. Thus, the future stock trend can be modeled by the function $F_{\hat{\theta}_T}$.

To train our paradigm, one feasible way is to first obtain the ``ground-truth" parameter at time $T$ by $(X_T, Y_T)$ with a first-order model. Then we minimize the gap between the ground-truth and the estimated parameter $\hat{\theta}_T$. However, here, note that the ``ground-truth'' parameter obtained by the first-order model is also an empirical estimation. Directly learning such ``ground-truth" would cause the error accumulation. Instead, we directly optimize the final prediction and the stock trend. The loss function $\mathcal{L}$ can be defined by 
\begin{equation}
    \mathcal{L} = (F_{\hat{\theta}_T}(X_{T}) - Y_{T})^2.
\end{equation}

% corresponding weight of \(s\)-day scale. $\alpha^s$ is also a trainable parameter, which is learned together with the other parameters. $\hat{\theta}_{T}$ is our guess for the prediction model at time step $T$.

% Since the empirically estimated parameter $\hat{\theta}$ within a certain margin of error cannot be regarded as the exact label of the second-order model, we design a unified framework to predict the stock price with the objective function as
%first, we apply the predicted parameters $\hat{\theta_T}$ on the predicted data to obtain the predicted stock price, then the square loss between the ground-truth and the predicted price are minimized by gradient descent. Now we formalize the objective function of the second-order learning paradigm as,

We display the whole framework of the second learning paradigm in Figure ~\ref{second_model} and formulate the process of the second-order sequential model in Algorithm \ref{alg:second-order_paradigm}.
% As the algorithm shows, 
% The second-order sequential model learns $F_{\theta}$ everyday to update the parameter collection, since today's parameter will be used as the input of the next-day's paradigm. This is different from the first-order learning paradigm which trains only once, then predicts on arbitrary datasets. 

\begin{algorithm}[t]
\caption{Stock prediction by second-order sequential model.} %算法的名字
 {\bf Input:} %算法的输入， \hspace*{0.02in}用来控制位置，同时利用 \\ 进行换行
Training set $D_{1}=\{(X_t, Y_{t+1})| t \in [0, T_a-1]\}$. \\
Testing set $D_{2}=\{(X_t, Y_{t+1})| t \in [T_a, T_b]\}$.\\
Time-scale set $S$. Episode number $E$. \\
Second-order sequential model $G_{\Phi}$ with $K$ time steps.\\
 {\bf Output:} %算法的结果输出
Stock trends $O$.

{\bf Training second-order sequential model $G$.}
\begin{algorithmic}[1]
\State Initialize $G_{\Phi}$;
\State Generate parameters $\Theta=\{\theta^s_t|s \in S, t \in [0, T_a-1] \}$ from $D_{1}$;
\State Construct $G$'s training set $\{(X^g_t, Y_{t+1})|t \in [0, T_a-1]\}$ where
$X_t^g = \{\theta^s_{t-ks}, X_t|s \in S, k \in [1, K]\}$;
%$D_3 = \{(\theta^s_{t-ks}, X_t|s \in S, k \in [1, K]), Y_{t+1}|t \in [0, T-1]\}$
%\State Construct $G$'s training inputs $I_t = \{\theta^s_{t-ks}, X_t|s \in S, k \in [1, K]\}$
\For{$e \leftarrow 1$ to $E$}
\State $\Phi \leftarrow \Phi - \nabla_{\Phi}(G(X_t^g) - Y_{t + 1})^2$;
\EndFor
\end{algorithmic}

{\bf Predicting by second-order sequential model $G$.}
\begin{algorithmic}[1]
\State Prediction results $O \leftarrow \emptyset$;
\For{$t \leftarrow T_a$ to $T_b$}
\State $X^g_{t} = \{\theta^s_{t-ks}, X_{t}|s \in S, k \in [1, K]\}$;
\State $\hat{Y}_{{t+1}} = G_{\Phi}(X^g_{t})$;
\State $O \leftarrow O \cup \{\hat{Y}_{t+1}\}$;
\State Generate parameters $\Theta_t = \{\theta^s_{t}|s \in S\}$;
\State $\Theta \leftarrow \Theta \cup \Theta_t$;
\EndFor
\State \Return $O$;
\end{algorithmic}
\label{alg:second-order_paradigm}
\end{algorithm}

\section{Experiments}
\subsection{Experimental Setup}

{\bf Data Set.} We evaluate our method on the real-world stock data of the Chinese market from 2013 to 2017 in daily frequency~\footnote{We collect daily stock price and volume data from http://xueqiu.com/ and https://finance.yahoo.com/}. 
There are more than 2000 stocks in total, covering the vast majority of Chinese stocks. In order to model the market trend, we filter out several ``bad'' stocks which  are under suspended trading status for more than 10\% of trading days. After that, there are totally 1246 stocks that are used in our experiments. Furthermore, we follow the previous study ~\cite{kakushadze2016101} to compute totally 101 trading indicators as the input of the first-order model.

In the following experiments, we use the stock data from 2013 to 2016 for training and validation while use the data of 2017 for testing. In order to validate the models in different market states, the training set and validation set are randomly extracted from the whole period from 2013 to 2016. Specifically, we randomly sample \(1/10\) data from this period as the validation set, while the other \(9/10\) as the training set. 

~\\
\noindent{\bf Compared Methods.} To evaluate the effectiveness of our models, we compare the following methods:

\begin{itemize}[leftmargin=*]
%{Linear Model as first-order learning paradigm (LM)
\item \textbf{First-order Learning Paradigm for Linear Model ($\mathrm{Lin}$)}: The method is the vanilla combination with inputs. It learns static model parameters on the training set, then predicts the future trend of stocks on the test data directly. 
\item \textbf{Rotation Learning for Linear Model with Window Size w }($\mathrm{RoT}$-w): This method keeps generating the new model by rotation using merely the recent data within a certain time window, where $w$ is the corresponding window size.
% and corresponds to the scale of second-order learning paradigm.
% It can track the dynamic parameter because of its daily update.
\item \textbf{Second-order Sequential Model with s-scale }($\mathrm{Sec}$-s): This approach is a special case of our proposed model, where we only use a single scale $s$. We introduce this special case to demonstrate the effectiveness of the multi-scale design.
\item \textbf{Multi-scale Second-order Sequential Model ($\mathrm{multi}$-$\mathrm{Sec}$)}: This is our proposed model which captures how the optimal prediction model evolves over time with multi-scale second-order patterns.
\end{itemize}
In this paper, we consider the time scale in days, for example, $\mathrm{Sec}$-1 denotes the second-order sequential model with 1-day scale. $\mathrm{RoT}$-60 uses 60-day window to train the model. Furthermore, $\mathrm{multi}$-$\mathrm{Sec}$ combines the patterns with respect to several scales, including 1-day, 5-day (1 week), 10-day (2 weeks) and 20-day (1 month) in this paper.

%chi, you are the best ,try your best to write the experiments!!! evaluation metrics VS backtesting, frist, EVAL -> top 50 strategy, no time sequence, so no investment. just estimate the top-k guess. why use top-k guess, no bottom-k? because we invest will 
~\\
{\bf Evaluation Metrics.} To compare the stock prediction methods, we evaluate the performance of top-$K$ stocks sorted by the predicted daily returns in descending order. We adopt the most widely used metrics, {\em Annualized Return} (AR) and {\em Sharpe Ratio} (SHR) to evaluate the performance of stock prediction, i.e., 
\begin{itemize}[leftmargin=*]

\item \textit{Annualized Return} (AR) is a common profit indicator in finance, calculated by the mean return of selected $K$ stocks in a \(l\)-day period to one year. Specifically, 
\begin{equation}
    AR@K = \frac{1}{K} \sum_{d \in [1, l]}\sum_{i \in \Lambda_d} r^i_d \times \frac{365}{l},
\end{equation}
where $\Lambda_d$ is the collection of selected top-$K$ stocks in the $d$-th day, and $r^i_d$ represents the return of stock $i$ in the $d$-th day.
\item \textit{Sharpe Ratio}~\cite{sharpe1966mutual} (SHR) is a risk-adjusted profit measure that computes the return per unit of deviation. In a formal definition, 
\begin{equation}
    %SHR = \frac{E[R_a-R_b]}{\sqrt{var[R_a]}},
    SHR@K = \frac{1}{l}\sum_{d \in [1, l]}\frac{\frac{1}{K}\sum_{i \in \Lambda_d}(r^i_d-\bar{r}_d)}{\sqrt{var(\{r^i_d|i \in \Lambda_d\})}},
\end{equation}
where \(\bar{r}_d\) is the average return of the market in the $d$-th day. Thus, SHR is positively related to the return and negatively related to the risk of a strategy.
\end{itemize}
To evaluate methods from various aspects, we respectively study the performance in top-$\{10, 20, 50\}$ strategies.

~\\
\noindent{\bf Hyperparameter Settings.} 
We employ the grid search to select the optimal hyperparameters regarding MSE on the validation sets for all methods. For LSTM parts of the models, we tune the number of LSTM cells within \{5, 10, 20\}, initialize the forget bias within \{0, 0.5, 1\} and tune the size of the hidden vector within \{64, 128\}.
%The L2 weight decay for $\theta$ is tuned within \{1e-2, 5e-3, 1e-3\}.

%\subsection{Overall Results}
\subsection{Results}

\begin{figure}[!t]
\setlength{\belowcaptionskip}{-10pt} 
\centering
\includegraphics[width=0.43\textwidth]{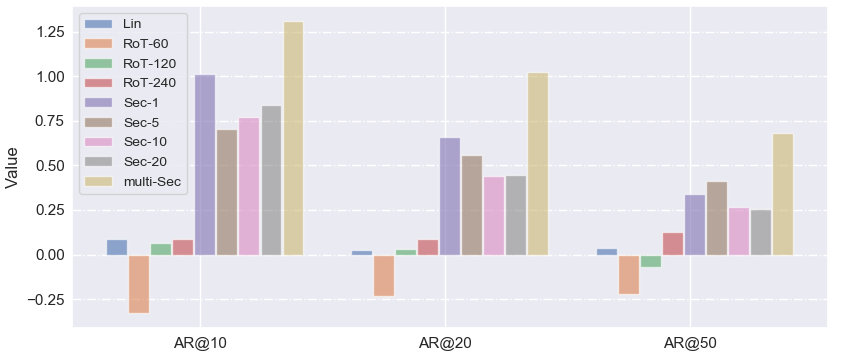}
\caption{The performance comparison on AR@10, 20, 50 among $\mathrm{Lin}$, $\mathrm{RoT}$, $\mathrm{Sec}$ and $\mathrm{multi}$-$\mathrm{Sec}$.}
\label{AR}
\end{figure}
%\vspace{-0.1in}
\begin{figure}[!t]
\setlength{\belowcaptionskip}{-10pt} 
\centering
\includegraphics[width=0.43\textwidth]{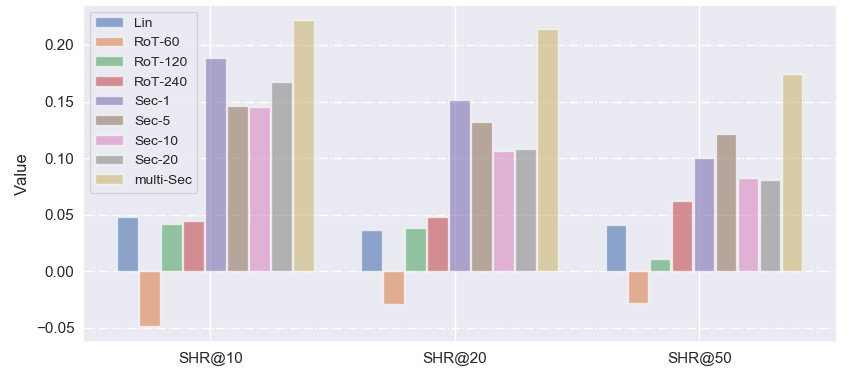}
\caption{The performance comparison on SHR @10, 20, 50 among $\mathrm{Lin}$, $\mathrm{RoT}$, $\mathrm{Sec}$ and $\mathrm{multi}$-$\mathrm{Sec}$.}

\label{SHR}
\end{figure}
%In this section, we compare the experiment results among LM, RLM, SSM, and \mathrm{multi-Sec}, then illustrate the effectiveness of the second-order learning paradigm.

{\bf Main Results.} Figure \ref{AR} and \ref{SHR} present the results among $\mathrm{Lin}$, $\mathrm{RoT}$, $\mathrm{Sec}$, and $\mathrm{multi}$-$\mathrm{Sec}$ on the test set. In general, $\mathrm{Sec}$ and $\mathrm{multi}$-$\mathrm{Sec}$ have significantly better performance than the other methods, which demonstrates that it is necessary to propose the second-order learning paradigm. Although $\mathrm{RoT}$ can update the first-order model dynamically, it is still much worse than our algorithm, which indicates that it is not enough to obtain a concrete prediction only by rotation learning. In terms of different scales of the proposed $\mathrm{Sec}$, $\mathrm{Sec}$-1 performs the best on the top-10 and top-20 investment while $\mathrm{Sec}$-5 brings the most profit on the top-50 investment, which states that different time scales brings different profit in the stock market. By combining different time scales, our proposed $\mathrm{multi}$-$\mathrm{Sec}$ achieves the best performance.

\noindent{\bf Rotation Learning Paradigm.} In Figures 4 and 5, $\mathrm{RoT}$-60 generates a money-losing investment, while $\mathrm{Lin}$, $\mathrm{Sec}$, and $\mathrm{multi}$-$\mathrm{Sec}$ can bring less or more profit. This is mainly due to that the rotation learning paradigm pays much attention on the recent data. However, since the stock market is highly dynamic, the method will suffer from the sudden distribution altering in the stock market. Enlarging the rotation window size alleviate this issue. Especially, in most cases, $\mathrm{RoT}$-240 outperforms $\mathrm{Lin}$ and $\mathrm{RoT}$ with the other window sizes.

% This can be caused by the following reasons: firstly, RLM updates the predictor within a small historical data, which overfits easily. Secondly, the predictor at each time step is trained by recent days, and the learned market state may be extremely different from the future one. Therefore, it is inadequate to merely consider the recent data. The second-order patterns should be paid attention in the real stock market. 
%Furthermore, RLM-1 obtains better performance than other RLMs. This illustrates 1-day second-order patterns make the daily stock prediction better.

%\subsection{Second-order Patterns in Multiple Scale}

\noindent{\bf Single-scale vs. Multi-scale.} As Figures \ref{AR} and \ref{SHR} show, $\mathrm{multi}$-$\mathrm{Sec}$ significantly outperforms the single scale models. It demonstrates that diverse information from the multi-scale market states is beneficial to the stock prediction. In addition, the more stocks are invested, the more advantages are generated by the multi-scale design: $\mathrm{multi}$-$\mathrm{Sec}$ is larger 0.0337, 0.0624 and 0.0742 than $\mathrm{Sec}$-1 on respectively SHR@$10, 20, 50$. This indicates that multi-scale information is especially useful to the diversified investment.
%It is because that the macroscale ignores a lot of noise in comparison to microscale. 

\noindent {\bf \boldmath$\alpha^s$ Value.} In order to study the contribution made by each scale, we print the magnitude of weight \(\alpha^s\) in each scale: 0.1357 on 1-day scale, 0.1393 on 5-day scale, 0.1353 on 10-day scale and 0.2290 on 20-day scale. There are a couple of observations from \(\alpha^s\) distribution: the 1-day, 5-day and 10-day scale have similar absolute weights, which indicates that the three scales contains similar information. In the meantime, the distinctly higher weight of 20-day scale implies that the 20-day scale brings very different information from the other scales, and is precious for stock prediction.

\noindent{\bf Case Studies.} To compare the single-scale and multi-scale design, Table \ref{case_study} shows the predicted weight of trading indicator MA$_{10}$ by $\mathrm{Sec}$ with different scales and $\mathrm{multi}$-$\mathrm{Sec}$. As the table shows, $\mathrm{multi}$-$\mathrm{Sec}$ and ground-truth have similar second-order trends with co-directional weights (-+-+-). This illustrates that our proposed $\mathrm{multi}$-$\mathrm{Sec}$ can model the reversal trend of second-order sequence which cannot be captured by rotation learning paradigm because it assumes the same data distribution between the recent data and the predicted data. Furthermore, $\mathrm{Sec}$ in distinct scales have different second-order patterns, for example, the trend of $\mathrm{Sec}$-1 is (down, up, up, down) from 2017/03/27 to 2017/03/31, while $\mathrm{Sec}$-20 corresponds to (up, down, down up).

\begin{table}[!tp]
\setlength{\belowcaptionskip}{-10pt} 
\centering
\caption{The predicted weight of indicator MA$_{10}$ in different scale.}
\small
\begin{tabular}{|c|c|c|c|c|c|c|c|c|c|}
\hline

   Date&Ground-truth&Sec-1&Sec-5\\
   \hline
   2017-03-27&-0.0226&0.0076&0.0486\\
   \hline
    2017-03-28&0.1858&0.0049&0.0478\\
    \hline
    2017-03-29&-0.0120&0.0070&0.0409\\
    \hline
    2017-03-30&0.0633&0.0307&0.0046\\
    \hline
    2017-03-31&-0.0254&0.0198&0.0288\\
    \hline
    Date&Sec-10&Sec-20&multi-Sec\\
   \hline
   2017-03-27&0.0363&0.0078&-0.0408\\
   \hline
    2017-03-28&0.0307&0.0398&0.0031\\
    \hline
    2017-03-29&0.0075&0.0334&-0.0196\\
    \hline
    2017-03-30&0.0088&0.0054&0.0083\\
    \hline
    2017-03-31&0.0346&0.0078&-0.0277\\

\hline
\end{tabular}
\label{case_study}
\end{table}

\subsection{Market Trading Simulation}
To further evaluate the effectiveness of our proposed models, we conduct the back-testing by simulating the stock trading for the test dataset. Our estimation strategy conducts trading in the daily frequency. 
Given a certain principal at the beginning of the back-testing, investors invest in the top-\(K\) stocks with the highest predicted return in each day. The selected stocks are held for one day. The cumulative profit without consideration of transaction cost will be invested in the next trading day.
We also calculate the average return on the stock market by evenly holding every stock as the baseline, indicating the overall market trend.

\begin{figure*}[!tp]
\centering
\includegraphics[width=1.0\textwidth]{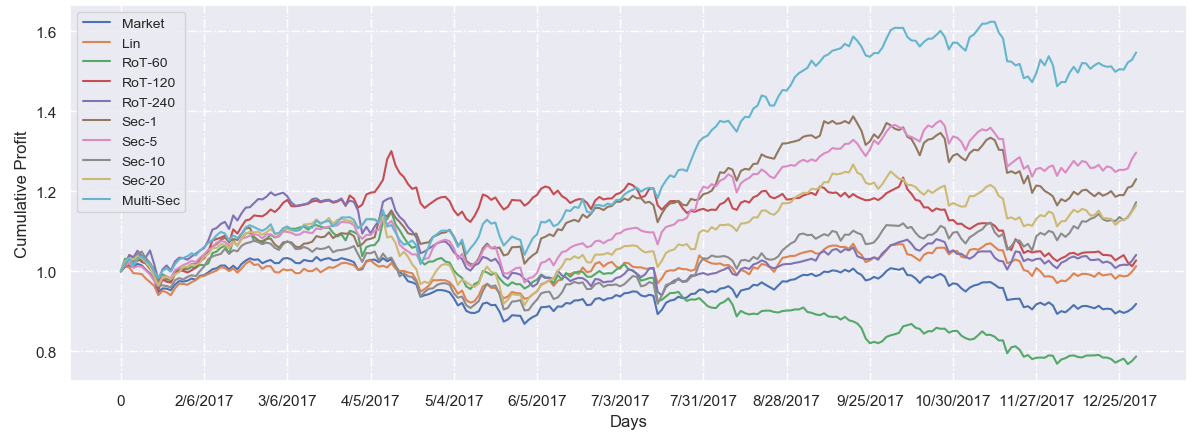}
\caption{The cumulative profit curve of differnt methods with the portfolio of chossing top 50 stocks.}
\label{backtesting}
\end{figure*}

Figure ~\ref{backtesting} shows the cumulative profit curve for each method with \(K\) as 50. As can be seen, our proposed second-order learning paradigm, $\mathrm{Sec}$ and $\mathrm{multi}$-$\mathrm{Sec}$, have the most profitable results over all baselines. In particular, $\mathrm{multi}$-$\mathrm{Sec}$ performs the best during a long period. Despite in the first half of 2017, rotation learning paradigm performs well and even achieve more profit than our algorithm, it loses a lot of money on the second half of 2017 due to the sudden distribution altering.
%However, our algorithm is much more stable.
% , even when the stock market is getting down before July. 
In the second half of 2017, much more profit can be brought by $\mathrm{multi}$-$\mathrm{Sec}$, because $\mathrm{multi}$-$\mathrm{Sec}$ considers both short-term and long-term market states while $\mathrm{Sec}$ merely models single time scale.
Furthermore, the performance of different time scale is alternating: $\mathrm{Sec}$-20 performs the best in March, while $\mathrm{Sec}$-5 generates the most profit after October. It indicates that the scale preference of the stock market is changing over time. In future work, we will consider to dynamically combine the multi-scale trends for more accurate prediction.

\section{Related Work}
In this section, we elaborate the related work for stock prediction in two parts: the first part is traditional methods including technical analysis and fundamental analysis, the second part is the machine learning techniques.
%Stock trend prediction has attracted many research efforts due to its decisive role in stock investment. Traditional approaches predict the future trend of stocks by technical analysis and fundamental analysis. With the rapid growth of machine learning techniques, many people take great efforts to apply various machine learning techniques for stock prediction.

Technical analysis deals with the time-series historical market data, such as trading price and volume, and make predictions based on that. Due to the noisy nature of the stock market, technical analysts not only use raw price and volume data, but also explore many sorts of technical indicators~\cite{colby1988encyclopedia}, which are mathematical transformations of price, volume and other inputs.
%These indicators are used to help predict the stock trend based on a series of chart patterns summarized by experienced investors such as well-known head and shoulders~\cite{osler1995head} and cup and handle~\cite{suh2004novel} etc. 
One major limitation of the technical analysis is that it is incapable of unveiling the rules that govern the dynamics of the market beyond price data. Fundamental analysis~\cite{abarbanell1997fundamental}, on the contrary, evaluates a stock in an attempt to assess its intrinsic value, by examining related economic, financial, and other qualitative and quantitative factors. 
Besides traditional technical/fundamental indicators, online information, such as news and forum, can also help people make better investment decision \cite{nassirtoussi2015text,zhou2016can}. 
%Besides traditional financial factors such as revenues, earnings, future growth, return on equity and profit margins, fundamental analysts also collect the stock and market information from online content, such as news and forum, to help them make better investment decision \cite{si2013exploiting,nassirtoussi2015text,zhou2016can}. 

Recently, machine learning techniques, which can automatically recognize the underlying patterns in the stock market with little human knowledge, have attracted many investors' attention. Substantial researcher have already tried various models with multiple input indicators for stock prediction, such as linear regression~\cite{bermingham2011using,mittal2012stock,izzah2017mobile}, decision tree~\cite{delen2013measuring,ballings2015evaluating} and neural networks~\cite{rather2015recurrent,ding2015deep,hafezi2015bat}. Among existing machine learning techniques, the linear model has good interpretability, while non-linear models can capture the complex patterns. With the development of deep learning, many works use the Recurrent Neural networks (RNNs) for stock prediction because they can model strong temporal dynamics of the stock market. Recent work obtains more competitive performance on RNNs, for example, Nelson \textit{et.al} built an LSTM network with a set of technical indicators as input to predict the stock trend \cite{nelson2017stock}. Zhang \textit{et.al} proposed SFM \cite{zhang2017stock} and applied it in the stock prediction task. Compared to LSTM, SFM decomposes the hidden states of memory cells into multiple frequency components, each of which models a particular frequency of latent trading patterns. The learned models by these machine learning methods characterize the underlying patterns of the stock market, and will be used in an arbitrary dataset constantly for future prediction.

No matter how complex existing models are, they are all designed based on the first-order i.d. assumption which assumes the stationary data distribution over time. However, due to the highly dynamic nature of the stock market, it is not adequate to predict the stock price on the strict first-order i.d. assumption.
%To address this challenge, we propose to relax it to second-order identical distribution assumption that considers the evolving data distribution. Furthermore, we design a second-order learning paradigm to learn the second-order patterns. Meanwhile, our proposed model also takes into account the multi-scale information for more accurate prediction.

\section {Conclusion}
In this paper, we address the dynamic and non-stationary property of the stock market, by introducing a second-order i.d. assumption. In contrast to existing methods that use the fixed model over time, we assume that the optimal prediction model is changing over time with certain patterns. Based on this assumption, we develop a second-order learning paradigm for capturing the second-order patterns. Furthermore, to presume more accurate prediction, the  proposed model can capture the evolving second-order pattern with respect to both micro-scale and macro-scale. In the end, extensive experiments on real-world Chinese stock market data demonstrate that our approach can result in a significant improvement.

In the future, we will extend our work to other first-order models. Due to lacking the clear economic meaning, it will be more challenging to model the second-order evolving pattern of non-linear model. 
In addition, considering the alternating performance of the different scale method in back-testing, we plan to dynamically combine the multi-scale information for more profit.

\bibliographystyle{named}
\small{
\bibliography{ijcai19}
}
\end{document}